\documentclass[aps,preprint,amsmath,amssymb,amsfonts,epsfig]{revtex4}
\usepackage{graphicx}
\usepackage{morefloats}
\begin{document}

\title{Search for anomalous quartic $ZZ\gamma\gamma$ couplings in photon-photon collisions }

\author{M. K\"{o}ksal}
\email[]{mkoksal@cumhuriyet.edu.tr} \affiliation{Department of Optical Engineering, Cumhuriyet University,
58140, Sivas, Turkey}
\author{V. Ar{\i}}
\email[]{Volkan.Ari@science.ankara.edu.tr} \affiliation{Department of Physics, Ankara University,
06100, Ankara, Turkey}
\author{A. Senol}
\email[]{senol_a@ibu.edu.tr} \affiliation{Department of Physics,
Abant Izzet Baysal University, 14280, Bolu, Turkey}

\begin{abstract}
The self-couplings of the electroweak gauge bosons are completely
specified by the non-Abelian gauge nature of the Standard Model
(SM). The direct study of these couplings provides a significant
opportunity to test the validity of the SM and the existence of new
physics beyond the SM up to the high energy scale. For this reason,
we investigate the potential of the processes
$\gamma\gamma\rightarrow ZZ$, $e^{-}\gamma\rightarrow
e^{-}\gamma^{*}\gamma \rightarrow e^{-}Z\, Z$ and $e^{+}e^{-}
\rightarrow e^{+}\gamma^{*} \gamma^{*} e^{-} \rightarrow e^{+}\, Z\,
Z\, e^{-}$ to examine the anomalous quartic couplings of
$ZZ\gamma\gamma$ vertex at the Compact Linear Collider (CLIC) with
center-of-mass energy $3$ TeV. We calculate $95\%$
confidence level sensitivities on the dimension-8 parameters with various values of the
integrated luminosity. We show that the best bounds on the anomalous $\frac{f_{M2}}{\Lambda^{4}}$, $\frac{f_{M3}}{\Lambda^{4}}$, $\frac{f_{T0}}{\Lambda^4}$ and $\frac{f_{T9}}{\Lambda^4}$
couplings arise from $\gamma\gamma\rightarrow ZZ$ process  among those three processes at center-of-mass energy of 3 TeV and integrated luminosity of $L_{int}=2000$ fb$^{-1}$ are found to be $[-3.30;3.30]\times 10 ^{-3}$ TeV$^{-4}$, $[-1.20;1.20]\times 10 ^{-2}$ TeV$^{-4}$, $[-3.40;3.40]\times 10 ^{-3}$ TeV$^{-4}$ and $[-1.80;1.80]\times 10 ^{-3}$ TeV$^{-4}$, respectively.
\end{abstract}
\maketitle

\section{Introduction}
The SM of particle physics has been tested with a lot of different
experiments for decades and it is proven to be extremely
successful. In addition, the discovery of all the particles predicted by the SM has been
completed together with the ultimate discovery of the approximately 125
GeV Higgs boson in 2012 at the Large Hadron Collider (LHC)
\cite{higgs1,higgs2}. However, we need a new physics beyond the SM
to find answers to some fundamental questions, such as the strong CP
problem, neutrino oscillations and matter - antimatter asymmetry in the universe.
The self-interactions of electroweak gauge bosons are important and more sensitive for new physics beyond the SM.
The structure of gauge boson self-interactions is completely determined by the non-Abelian $SU(2)_{L}\otimes U(1)_{Y}$ gauge symmetry in the SM.  Contributions to these interactions, beyond those coming from the SM, will be a supporting evidence of probable new physics. It can be examined in a
model independent way via the effective Lagrangian approach. Such an
approach is parameterized by high-dimensional operators which
induce anomalous quartic gauge couplings that modify the
interactions between the electroweak gauge bosons. 

In writing effective operators associated to genuinely quartic couplings we employ the formalism of Refs. \cite{Belanger:1992qh, Belanger:1992qi}. Imposing global $SU(2)_L $ symmetry and local $U(1)_Y$ symmetry, dimension-6 effective Lagrangian for the $ZZ\gamma\gamma$  coupling is given by
\begin{eqnarray}
\mathcal{L}=\mathcal{L}_{0}+\mathcal{L}_{c},
\end{eqnarray}
\begin{eqnarray}
\mathcal{L}_{0}=\frac{-\pi\alpha}{4}\frac{a_{0}}{\Lambda^{2}}F_{\mu \nu}F^{\mu \nu} W_{ \alpha}^{(i)} W^{(i) \alpha},
\end{eqnarray}
\begin{eqnarray}
\mathcal{L}_{c}=\frac{-\pi\alpha}{4}\frac{a_{c}}{\Lambda^{2}}F_{\mu \alpha}F^{\mu \beta} W^{(i)\alpha} W_{\beta}^{(i)}
\end{eqnarray}
where $F_{\mu \nu}=\partial_{\mu}A_{\nu}-\partial_{\nu}A_{\mu}$ is
the tensor for electromagnetic field tensor, and $a_{0,c}$ are the
dimensionless anomalous quartic coupling constants, $\Lambda$ is a
mass-dimension parameter associated with the scale of new physics.

The anomalous quartic gauge couplings come out also from dimension-8 operators. There are three classes of operators containing either covariant derivatives of Higgs doublet ($D_\mu \Phi$) only,  or two field strength tensors and two $D_\mu \Phi$, or field strength tensors only. The first class operators contain anomalous quartic gauge couplings involving only massive vector boson. We will not examine them since these operators  contain only quartic $W^+W^-W^+W^-$, $W^+W^-ZZ$ and $ZZZZ$ interactions. In the second class, eight anomalous quartic gauge boson couplings are given by \cite{yuk,yuk1,yuk2}
\begin{eqnarray}
\mathcal{L}_{M0}&=&\frac{f_{M0}}{\Lambda^{4}}Tr[W_{\mu\nu}W^{\mu\nu}]\times[(D_{\beta}\Phi)^{\dagger}D^{\beta}\Phi],\\
\mathcal{L}_{M1}&=&\frac{f_{M1}}{\Lambda^{4}}Tr[W_{\mu\nu}W^{\nu\beta}]\times[(D_{\beta}\Phi)^{\dagger}D^{\mu}\Phi],\\
\mathcal{L}_{M2}&=&\frac{f_{M2}}{\Lambda^{4}}[B_{\mu\nu}B^{\mu\nu}]\times[(D_{\beta}\Phi)^{\dagger}D^{\beta}\Phi],\\
\mathcal{L}_{M3}&=&\frac{f_{M3}}{\Lambda^{4}}[B_{\mu\nu}B^{\nu\beta}]\times[(D_{\beta}\Phi)^{\dagger}D^{\mu}\Phi],\\
\mathcal{L}_{M4}&=&\frac{f_{M4}}{\Lambda^{4}}[(D_{\mu}\Phi)^{\dagger}W_{\beta\nu} D^{\mu}\Phi]\times B^{\beta\nu},\\
\mathcal{L}_{M5}&=&\frac{f_{M5}}{\Lambda^{4}}[(D_{\mu}\Phi)^{\dagger}W_{\beta\nu} D^{\nu}\Phi]\times B^{\beta\mu},\\
\mathcal{L}_{M6}&=&\frac{f_{M6}}{\Lambda^{4}}[(D_{\mu}\Phi)^{\dagger}W_{\beta\nu}W^{\beta\nu} D^{\mu}\Phi],\\
\mathcal{L}_{M7}&=&\frac{f_{M7}}{\Lambda^{4}}[(D_{\mu}\Phi)^{\dagger}W_{\beta\nu}W^{\beta\mu} D^{\nu}\Phi].
\end{eqnarray}
where the field strength tensor of the $SU(2)$ ($W_{\mu\nu}$) and $U(1)$ ($B_{\mu\nu}$) are given by
\begin{eqnarray}
W_{\mu\nu}&=&\frac{i}{2} g \tau^i ( \partial_{\mu} W_{\nu}^i-\partial_{\nu}W_{\mu}^i+g \epsilon_{ijk} W_{\mu}^j W_{\nu}^k)\nonumber\\
B_{\mu\nu}&=&\frac{i}{2}g'(\partial_{\mu}B_{\nu}-\partial_{\nu}B_{\mu}).
\end{eqnarray}
Here, $\tau^i (i=1,2,3)$ are the $SU(2)$ generators, $g=e/sin\theta_W$, $g'=g/cos \theta_W$, $e$ is the unit of electric charge and $\theta_W$ is the Weinberg angle.  The dimension-6 operators can be expressed simply in terms of dimension-8 operators due to their similar Lorentz structures.  The following expressions show the relations between the $f_{M_i}$ couplings for the  $ZZ\gamma\gamma$ vertex and $a_0$ and the $a_c$ couplings,  needed to compare with the LEP results;
\begin{eqnarray}
\frac{f_{M0}}{\Lambda^4} =\frac{a_0}{\Lambda^2}\frac{1}{g^2v^2}  &\textrm{ and} & \frac{f_{M1}}{\Lambda^4} =-\frac{a_c}{\Lambda^2}\frac{1}{g^2v^2} \\ \nonumber\\
\frac{f_{M2}}{\Lambda^4} =\frac{a_0}{\Lambda^2}\frac{2}{g^2v^2}  & \textrm{and} & \frac{f_{M3}}{\Lambda^4} =-\frac{a_c}{\Lambda^2}\frac{2}{g^2v^2} \\ \nonumber\\
\frac{f_{M4}}{\Lambda^4} =\pm\frac{a_0}{\Lambda^2}\frac{1}{g^2v^2}  & \textrm{and} & \frac{f_{M5}}{\Lambda^4} =\pm\frac{a_c}{\Lambda^2}\frac{2}{g^2v^2}\\ \nonumber\\
\frac{f_{M6}}{\Lambda^4} =\frac{a_0}{\Lambda^2}\frac{2}{g^2v^2}  & \textrm{and} & \frac{f_{M7}}{\Lambda^4} =\frac{a_c}{\Lambda^2}\frac{2}{g^2v^2} 
\end{eqnarray}

The operators containing four field strength tensors lead to quartic anomalous couplings are as follows
\begin{eqnarray}
\mathcal{L}_{T0}&=&\frac{f_{T0}}{\Lambda^4}\textrm{Tr}[W_{\mu\nu}W^{\mu\nu}]\times \textrm{Tr}[W_{\alpha\beta}W^{\alpha\beta}]\\
\mathcal{L}_{T1}&=&\frac{f_{T1}}{\Lambda^4}\textrm{Tr}[W_{\alpha\nu}W^{\mu\beta}]\times \textrm{Tr}[W_{\mu\beta}W^{\alpha\nu}]\\
\mathcal{L}_{T2}&=&\frac{f_{T2}}{\Lambda^4}\textrm{Tr}[W_{\alpha\mu}W^{\mu\beta}]\times \textrm{Tr}[W_{\beta\nu}W^{\nu\alpha}]\\
\mathcal{L}_{T5}&=&\frac{f_{T5}}{\Lambda^4}\textrm{Tr}[W_{\mu\nu}W^{\mu\nu}]\times B_{\alpha\beta}B^{\alpha\beta}\\
\mathcal{L}_{T6}&=&\frac{f_{T6}}{\Lambda^4}\textrm{Tr}[W_{\alpha\nu}W^{\mu\beta}]\times B_{\mu\beta}B^{\alpha\nu}\\
\mathcal{L}_{T7}&=&\frac{f_{T7}}{\Lambda^4}\textrm{Tr}[W_{\alpha\mu}W^{\mu\beta}]\times B_{\beta\nu}B^{\nu\alpha}\\
\mathcal{L}_{T8}&=&\frac{f_{T8}}{\Lambda^4}[B_{\mu\nu}B^{\mu\nu}B_{\alpha\beta}B^{\alpha\beta}]\\
\mathcal{L}_{T9}&=&\frac{f_{T9}}{\Lambda^4}[B_{\alpha\mu}B^{\mu\beta}B_{\beta\nu}B^{\nu\alpha}]
\label{d8}
\end{eqnarray}
where $f_{T0}$, $f_{T1}$, $f_{T2}$, $f_{T5}$, $f_{T6}$, $f_{T7}$, $f_{T8}$ and $f_{T9}$ are dimensionless parameters which have no dimensions-6 analogue.

The experimental 95 \% C.L. bounds on dimension-6 $ZZ\gamma\gamma$ couplings at the LEP by OPAL collaboration through the process $e^{+}e^{-}\rightarrow Z \gamma \gamma \rightarrow q \bar{q} \gamma \gamma$ are \cite{Abbiendi:2004bf}

\begin{eqnarray}
-0.007 ~\textmd{GeV}^{-2}<\frac{a_{0}}{\Lambda^{2}}<0.023 ~\textmd{GeV}^{-2},\\
-0.029 ~\textmd{GeV}^{-2}<\frac{a_{c}}{\Lambda^{2}}<0.029 ~\textmd{GeV}^{-2}.
\end{eqnarray}

The 95 \% C.L. one-dimensional bounds on dimension-8 parameters at the LHC by ATLAS collaboration through $q \bar{q}\rightarrow Z \gamma \gamma$ with an integrated luminosity of 20.3 fb$^{-1}$ at $\sqrt s $= 8 TeV \cite{atlas} are
\begin{eqnarray}
-1.6\times 10^{4}~\textmd{TeV}^{-4}<\frac{f_{M2}}{\Lambda^{4}}<1.6\times 10^{4}~\textmd{TeV}^{-4},\\
-2.9\times 10^{4}~\textmd{TeV}^{-4}<\frac{f_{M3}}{\Lambda^{4}}<2.7\times 10^{4}~\textmd{TeV}^{-4},\\
-0.86\times 10^{2}~\textmd{TeV}^{-4}<\frac{f_{T0}}{\Lambda^{4}}<1.03\times 10^{2}~\textmd{TeV}^{-4},\\
-0.69\times 10^{3}~\textmd{TeV}^{-4}<\frac{f_{T5}}{\Lambda^{4}}<0.68\times 10^{3}~\textmd{TeV}^{-4},\\
-0.74\times 10^{4}~\textmd{TeV}^{-4}<\frac{f_{T9}}{\Lambda^{4}}<0.74\times 10^{4}~\textmd{TeV}^{-4}.
\end{eqnarray}

In the literature, the anomalous quartic $ZZ\gamma\gamma$ couplings have been performed with Monte-Carlo studies  at the linear $e^{+}e^{-}$ colliders via  the processes $e^{+}e^{-}\rightarrow Z \gamma \gamma$ \cite{Stirling:1999ek,Gutierrez-Rodriguez:2013eya}, $e^{+}e^{-}\rightarrow Z Z \gamma$ \cite{Belanger:1999aw}, $e^{+}e^{-}\rightarrow q \bar{q} \gamma \gamma$ \cite{Montagna:2001ej}, $e^{+}e^{-}\rightarrow e^{+}\gamma^{*}e^{-} \rightarrow e^{+} Z Z e^{-}$ \cite{murat}, $e \gamma\rightarrow Z Z e$ \cite{Eboli:1993wg},
$e\gamma\rightarrow Z \gamma e$ \cite{Atag:2007ct}, $e^{+}e^{-}\rightarrow e^{+}\gamma^{*}\gamma^{*}e^{-} \rightarrow e^{+} Z Z e^{-}$ \cite{murat} and $\gamma\gamma\rightarrow WWZ$ \cite{Eboli:1995gv}.
For the hadron colliders, studies have been done on anomalous quartic $ZZ\gamma\gamma$ couplings  via the processes $p\,\bar{p}\rightarrow Z \gamma \gamma $ \cite{Dervan:1999as}, $p\, p\, (\bar{p})\rightarrow \gamma \gamma \ell \ell$ \cite{Eboli:2000ad}, $pp\rightarrow p \gamma^{*} \gamma^{*} p \rightarrow p Z Z p$   \cite{Chapon:2009hh,Pierzchala:2008xc,Gupta:2011be,deFavereaudeJeneret:2009db}, $pp\rightarrow p \gamma^{*} p \rightarrow p Z Z q X$ \cite{Senol:2013lca}, $pp\rightarrow p \gamma^{*} p \rightarrow p \gamma Z q X$ \cite{Sahin:2012mz}, $pp\rightarrow qq \gamma \ell \ell$ \cite{Eboli:2003nq}.
\section {Photon colliders}
The LHC may not be an ideal platform to study new physics beyond the SM because of remnants arising from the strong interactions. On the other hand, the linear colliders usually supply a cleaner environment with respect to the hadron colliders. The CLIC is one of the most popular linear collider designs, and it will operate in three different centre-of-mass energy stages. Probable operating scenarios of  CLIC are planned with an integrated luminosity of $500$ fb$^{-1}$ at $0.35$ TeV, $1500$ fb$^{-1}$ at $1.4$ TeV and $2000$ fb$^{-1}$
at $3$ TeV collision energy \cite{clic}. Having high luminosity and energy is extremely significant in terms of new physics research.
Particularly, the anomalous quartic gauge couplings are described by means of high-dimensional effective
Lagrangians which have very strong energy dependences. For this reason, the sensitivity to the anomalous couplings increases with energy much faster than the sensitivity to the SM ones, and they can be measured with better precision. Also, the $e^{-}e^{+}$ colliders are more likely to produce three or more
massive gauge bosons in the final states of the studying processes. As a result, these colliders will provide an occasion to investigate the anomalous quartic gauge boson couplings.

The expected design of the future linear collider will include operation also in $e \gamma$
and  $\gamma \gamma$ modes. In $e \gamma$
and  $\gamma \gamma$ processes, real photon beams can be generated by
converting the incoming $e^{-}$ and $e^{+}$ beams into photon beams through the Compton backscattering
mechanism. The maximum collision energy is expected to be 80\% for  $\gamma \gamma$ collision and  90\% for $e \gamma$ collision of the original $e^+e^-$ collision energy. However, the expected luminosities are 15\% for $\gamma \gamma$ collision and 39\% for $ e \gamma$ collision of the $e^+e^-$ luminosities \cite{Telnov:2016lzw}. Also when using directly the lepton beams, quasi-real photons will be radiated at the interaction allowing for processes like  $e \gamma^{*}$, $\gamma \gamma^{*}$ and $\gamma^{*} \gamma^{*}$ to occur \cite{g1,g2,g3,g4,g5,g55}. Alternatively, a $\gamma^{*}$ photon emitted from either of the incoming leptons can interact with a laser photon backscattered from the other lepton beam, and the subprocess $\gamma\gamma^* \to ZZ$ can take place.  Hence, we calculate the process $e \gamma\rightarrow e \gamma^{*} \gamma \rightarrow e Z Z $ by integrating the cross section for the subprocess $\gamma\gamma^* \to ZZ$ over the  $\gamma^{*}$ flux. Furthermore, $\gamma^{*}$ photons emitted from both lepton beams can collide with each other and the subprocess $\gamma^*\gamma^* \to ZZ$ can be produced, and the cross section for the full process $e e\rightarrow e \gamma^{*} \gamma^{*} e\rightarrow e Z Z e$ is calculated by integrating the cross section for the subprocess $\gamma^*\gamma^* \to ZZ$ over both $\gamma^{*}$ fluxes. The quasi-real $\gamma^{*}$ flux in $\gamma \gamma^{*}$ and $\gamma^{*} \gamma^{*}$ collisions is defined by the Weizsacker-Williams approximation (WWA). In the WWA, the electro-production processes includes a small angle of charged particle scattering. The virtuality of $\gamma^{*}$ photons emitted by the scattering particle is very small. Hence, they are supposed to be almost real.  There is a possibility to reduce the process of electro-production to the photo-production described by the following photon spectrum \cite{g4};
\begin{eqnarray}
f(x,Q_{max})=\frac{\alpha}{2\pi} \Big\{[1+(1-x)^2]\log\frac{Q_{max}}{Q_{min}} -2m_e x^2[\frac{1}{Q_{min}}-\frac{1}{Q_{max}}]\Big\}
\end{eqnarray}
where $m_e$ is mass of the scattering particle, $Q_{min}=m_e^2 x/(1-x)$ and $x=E_{\gamma}/E$. $E_{\gamma}$ and $E$ are energy of photon and energy of scattered electron (positron), respectively. Many examples of investigation of possible new physics beyond the SM through photon-induced processes using the WWA are available in the literature \cite{g6,g7,g8,g9,g10, Sahin:2011yv,g11,g12,g13,g14,g15,g16,g17,g18,g19,g20,g21,g22,g23,g24,g25,g26,g27,g28,g29,g299,g30,g31}.

\section{ $ZZ$ production at $\gamma\gamma$, $\gamma^*\gamma^*$ and $\gamma^*\gamma$ collisions}
\begin{figure}[htbp]
\includegraphics{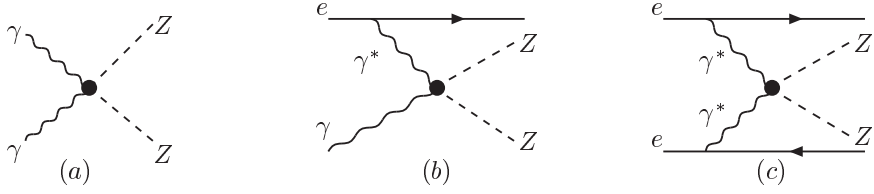}
\caption{Tree-level Feynman diagrams for the processes  (a) $\gamma\gamma \to ZZ$ (b)  $e^{-}\gamma\rightarrow e^{-}\gamma^{*}\gamma \rightarrow e^{-}Z\, Z$ (c) $e^{+}\gamma^{*} \gamma^{*} e^{-} \rightarrow e^{+}\, Z\, Z\, e^{-}$. }
\label{feyman}
\end{figure}

In this section we will display the differential cross sections by considering the contributions of all three types of collisions separately, $\gamma\gamma$, $\gamma^*\gamma^*$  and $\gamma^*\gamma$, for the $ZZ$ productions through the process $\gamma\gamma \to ZZ$ and the subprocesses $\gamma^*\gamma^* \to ZZ$ and $\gamma\gamma^* \to ZZ$.  The representative leading order Feynman diagrams of these process are given in Fig. \ref{feyman}. The dimension-8 anomalous interaction vertices in Eqs.  (4)-(11) and  Eqs. (17)-(24) are implemented in FeynRules \cite{Alloul:2013bka} and passed to MadGraph 5 \cite{Alwall:2014hca} framework by means of the UFO model \cite{Degrande:2011ua}.
\subsection{$\gamma\gamma$ collision}
\begin{figure}[htbp]
\includegraphics [width=0.8\columnwidth]  {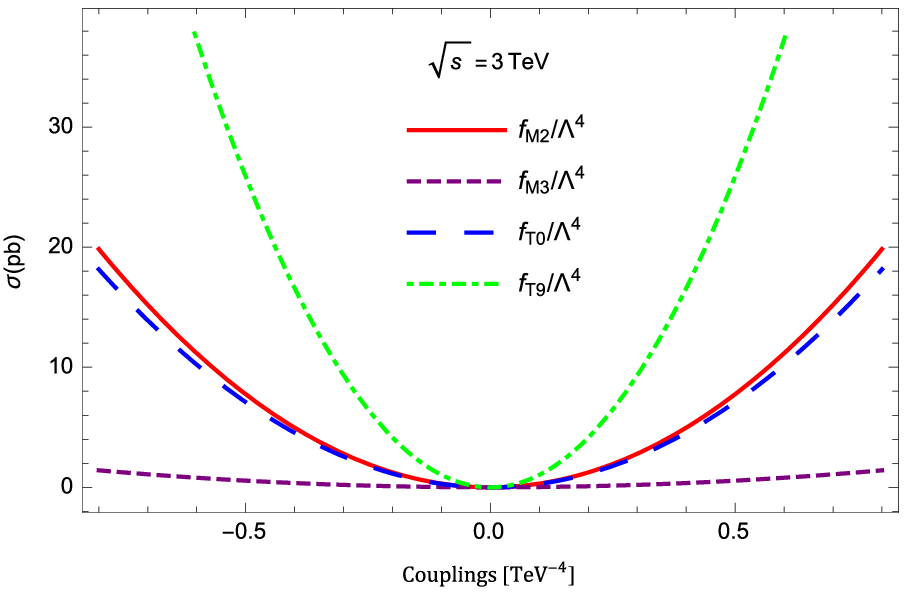}
\caption{The total cross sections as function of anomalous $f_{M2}/\Lambda^4$, $f_{M3}/\Lambda^4$, $f_{T0}/\Lambda^4$ and $f_{T9}/\Lambda^4$ couplings for the process $\gamma\gamma \to ZZ$ at  CLIC with $\sqrt s=3$ TeV.}
\label{Fig.1}
\end{figure}

The total cross section for the process $\gamma\gamma\to ZZ $ has been calculated by using real
photon spectrum produced by Compton backscattering of laser beam off the high energy electron beam. We show the total cross section of the process $\gamma\gamma \to ZZ$ depending on the dimension-8 anomalous couplings $f_{M2}/\Lambda^4$, $f_{M3}/\Lambda^4$, $f_{T0}/\Lambda^4$ and $f_{T9}/\Lambda^4$ for $\sqrt s$= 3 TeV in Fig. \ref{Fig.1}. In addition to these, the total cross sections as function of anomalous quartic couplings assuming  $\Lambda$=1 TeV are given in Table \ref{tab1}. In these figures, the cross sections depending on the anomalous quartic gauge couplings were obtained by varying only one of the anomalous couplings at a time while the others were fixed to zero. From these figures we can see that, the contribution comes from $f_{T9}/\Lambda^4$ coupling to the cross section grows rapidly while $f_{M2}/\Lambda^4$, $f_{M3}/\Lambda^4$ and $f_{T0}/\Lambda^4$ couplings are slowly varying. Hence the bounds on $f_{T9}/\Lambda^4$ coupling are expected to be more sensitive in accordance with $f_{M2}/\Lambda^4$, $f_{M3}/\Lambda^4$ and $f_{T0}/\Lambda^4$. Similarly, sensitivities on $f_{M2}/\Lambda^4$ and $f_{T0}/\Lambda^4$ couplings are expected to be more restrictive than sensitivities on $f_{M3}/\Lambda^4$.
\subsection{$\gamma^*\gamma^*$ collision}
\begin{figure}[htbp]
\includegraphics [width=0.8\columnwidth]  {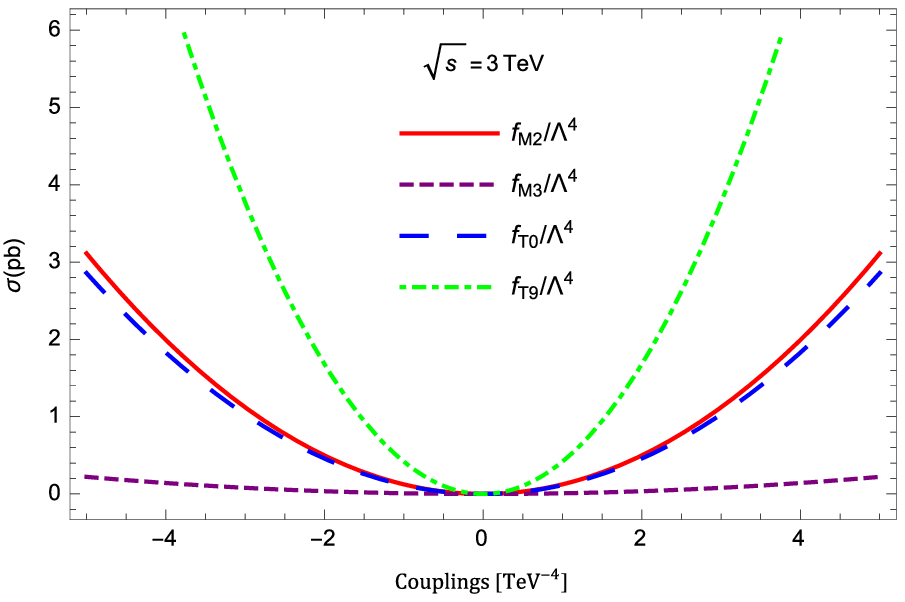}
\caption{The total cross sections as function of anomalous $f_{M2}/\Lambda^4$, $f_{M3}/\Lambda^4$, $f_{T0}/\Lambda^4$ and $f_{T9}/\Lambda^4$ couplings for the process $e^{+}e^{-}
\rightarrow e^{+}\gamma^{*} \gamma^{*} e^{-} \rightarrow e^{+}\, Z\,
Z\, e^{-}$ at CLIC with $\sqrt s=3$ TeV.}
\label{Fig.4}
\end{figure}

The $\gamma^*\gamma^* \to ZZ$  is generated via the quasi-real photons emitted from both lepton beams collision with each other, and participates as a subprocess in the main process $e^- e^+\rightarrow e^- \gamma^{*} \gamma^{*} e^+\rightarrow e^- Z Z e^+$. When calculating the total cross sections for this process, we take into account the equivalent photon approximation structure
function using the improved Weizsaecker-Williams formula which is embedded in MadGraph. The total cross sections of the process as a function of  $f_{M2}/\Lambda^4$, $f_{M3}/\Lambda^4$, $f_{T0}/\Lambda^4$ and $f_{T9}/\Lambda^4$ for $\sqrt s$= 3 TeV are given in Fig. \ref{Fig.4} and tabulated in
Table \ref{tab1}  assuming $\Lambda=1$ TeV.

\subsection{$\gamma\gamma^*$ collision}
\begin{figure}[htbp]
\includegraphics [width=0.8\columnwidth]  {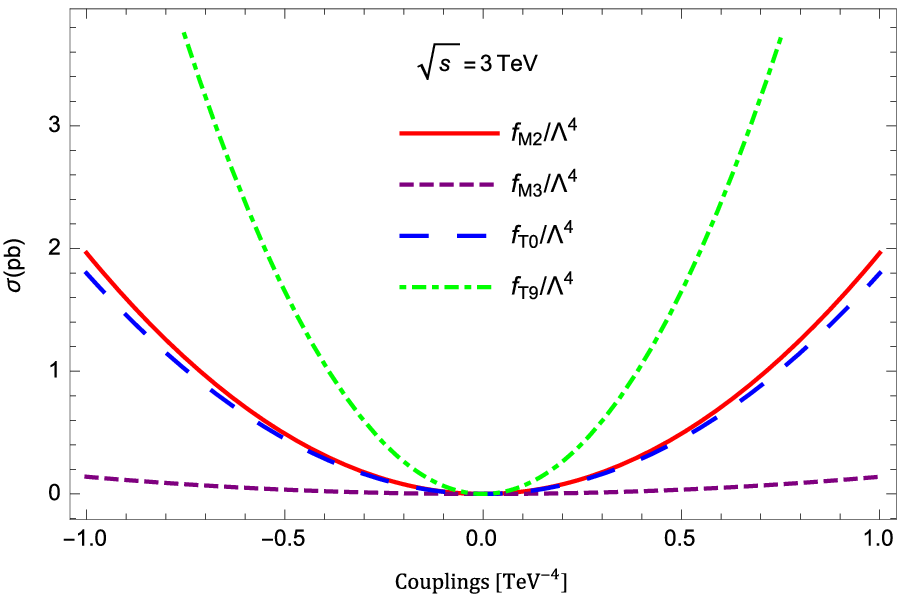}
\caption{The total cross sections as function of anomalous $f_{M2}/\Lambda^4$, $f_{M3}/\Lambda^4$, $f_{T0}/\Lambda^4$ and $f_{T9}/\Lambda^4$ couplings for the process $e^{-}\gamma\rightarrow
e^{-}\gamma^{*}\gamma \rightarrow e^{-}Z\, Z$ at CLIC with $\sqrt s=3$ TeV.}
\label{Fig.7}
\end{figure}

One of the operating mode of the conventional $e^+e^-$ machine is the $e\gamma$ mode. This mode includes $\gamma\gamma^*$ collision of a Weisaczker-Willams photon ($\gamma^{*}$) emitted from the incoming leptons and the laser backscattered photon ($\gamma$). Thus, the reaction $\gamma\gamma^* \to Z Z$ participates  as a subprocess in the main process $e^{-}\gamma\rightarrow
e^{-}\gamma^{*}\gamma \rightarrow e^{-}Z\, Z$. In Fig. \ref{Fig.7}, we plot the total cross section of the process $e^{-}\gamma\rightarrow e^{-}\gamma^{*}\gamma \rightarrow e^{-}Z\, Z$ as a function of dimension-8 couplings for $\sqrt s$= 3 TeV. Also, the total cross sections as function of anomalous quartic couplings assuming  $\Lambda$=1 TeV are given in Table \ref{tab1}.

 \section{Bounds on anomalous quartic couplings}
\begin{figure}[htbp]
\includegraphics [width=0.8\columnwidth] {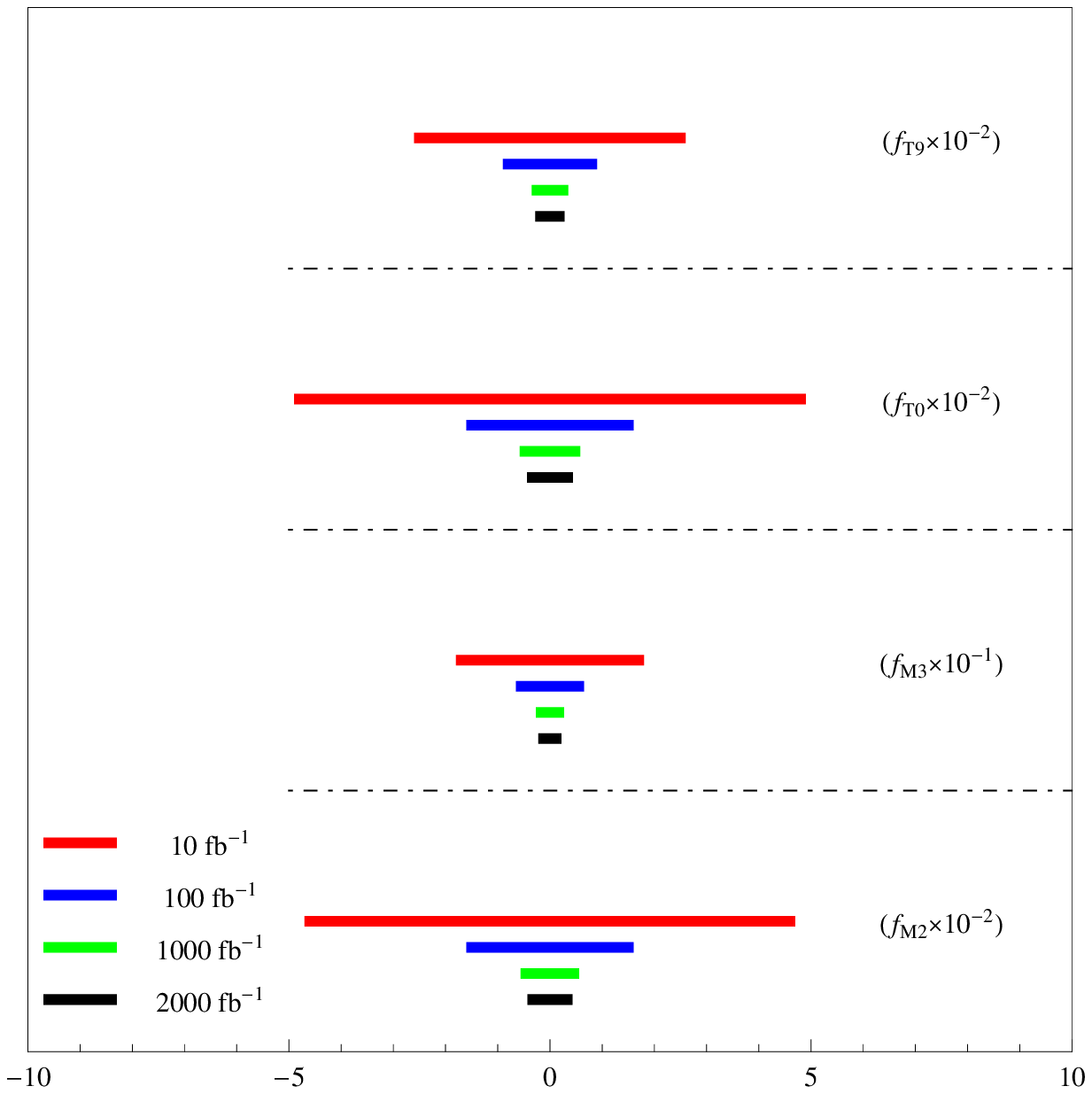}
\caption{$95\%$ C.L. sensitivity bounds of $f_{M2}/\Lambda^4$, $f_{M3}/\Lambda^4$, $f_{T0}/\Lambda^4$ and $f_{T9}/\Lambda^4$ couplings for various values of integrated luminosities and $\sqrt s=3$ TeV through the process $\gamma\gamma \to ZZ$.}
\label{Fig.10}
\end{figure}

\begin{figure}[htbp]
\includegraphics   [width=0.8\columnwidth] {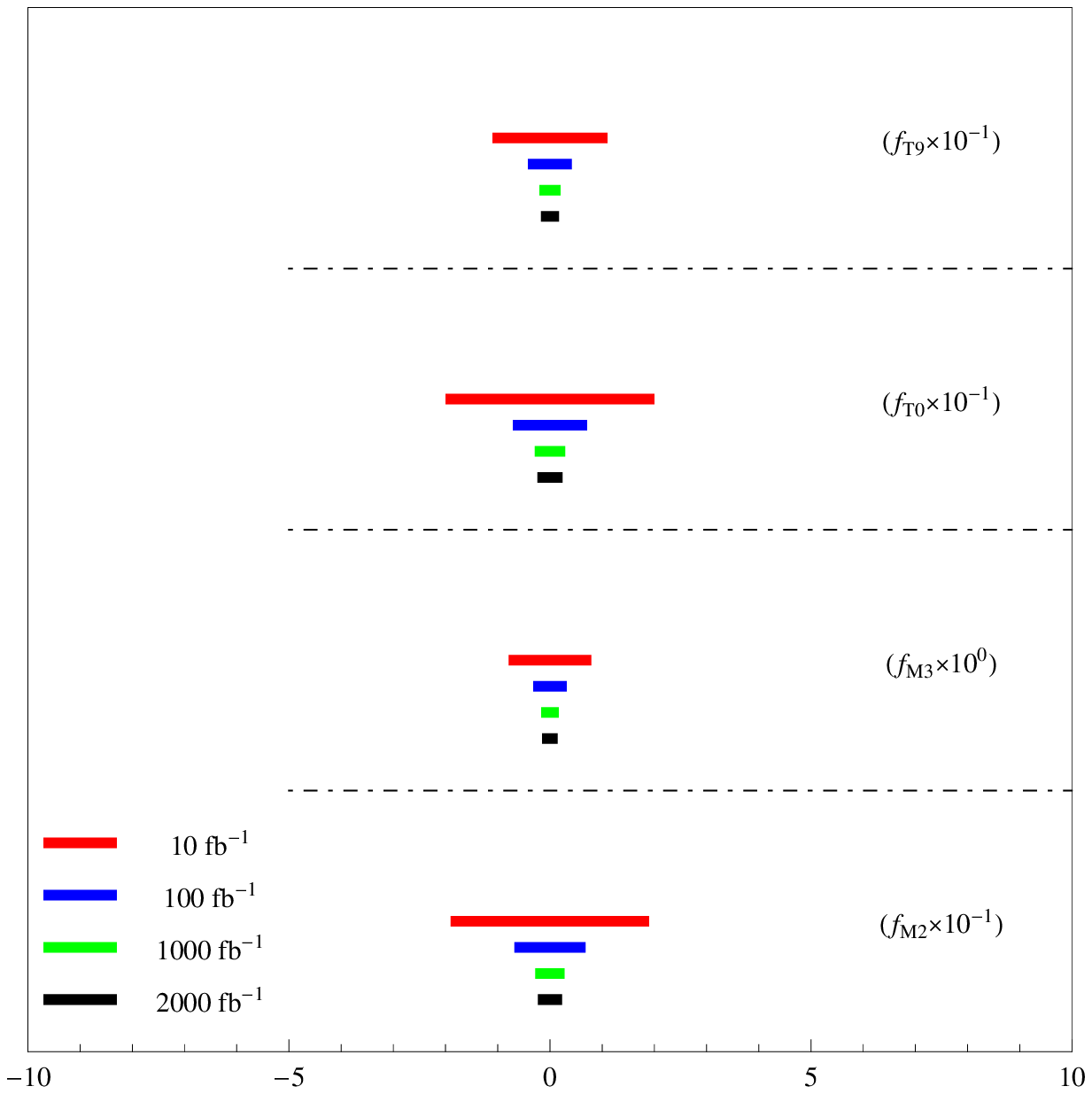}
\caption{$95\%$ C.L. sensitivity bounds of $f_{M2}/\Lambda^4$, $f_{M3}/\Lambda^4$, $f_{T0}/\Lambda^4$ and $f_{T9}/\Lambda^4$  couplings for various values of integrated luminosities and $\sqrt s=3$ TeV through the process $e^{-}\gamma\rightarrow
e^{-}\gamma^{*}\gamma \rightarrow e^{-}Z\, Z$.}
\label{Fig.14}
\end{figure}

\begin{figure}[htbp]
\includegraphics   [width=0.8\columnwidth]{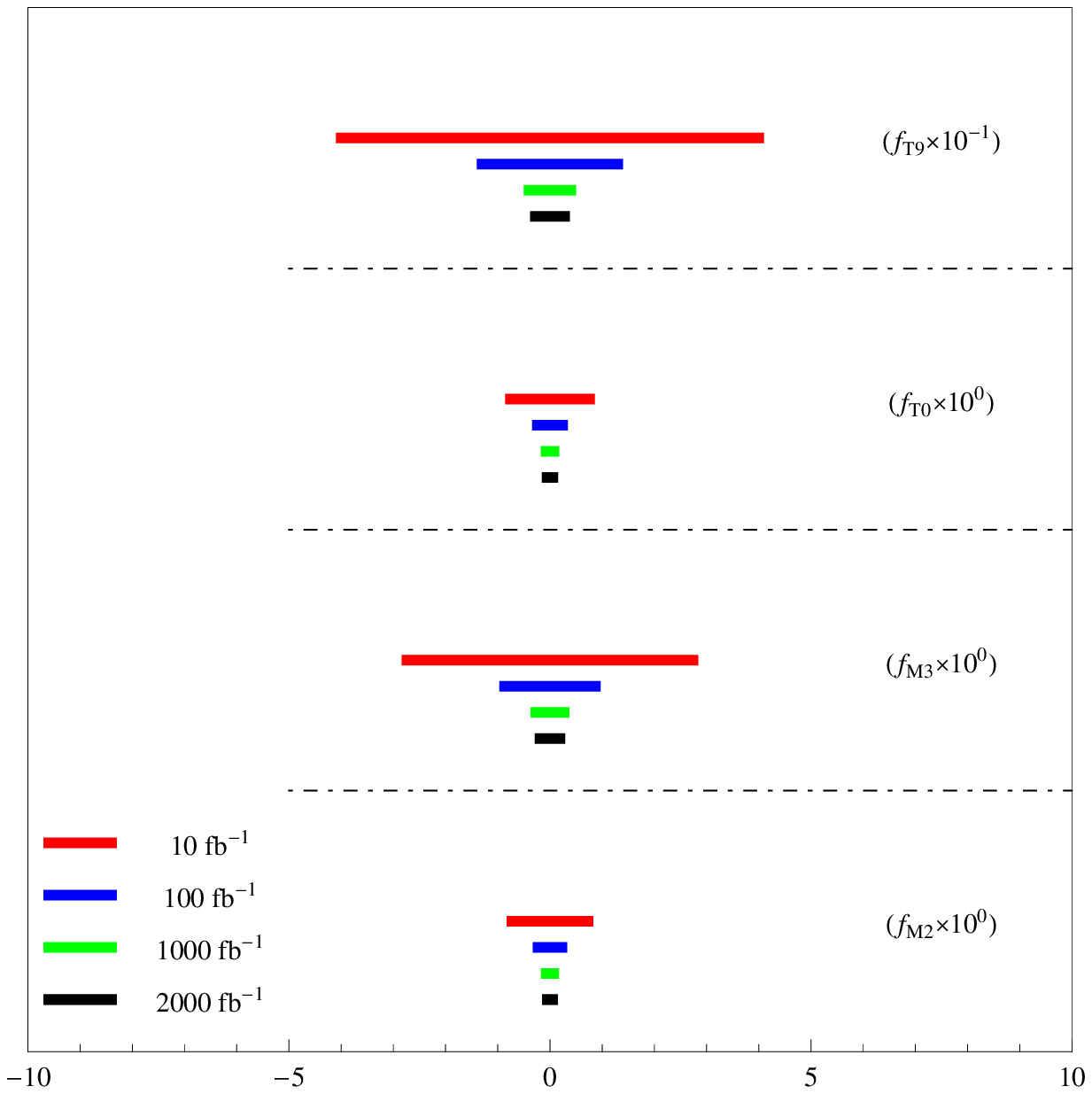}
\caption{$95\%$ C.L. sensitivity bounds of $f_{M2}/\Lambda^4$, $f_{M3}/\Lambda^4$, $f_{T0}/\Lambda^4$ and $f_{T9}/\Lambda^4$ couplings for various values of integrated luminosities and $\sqrt s=3$ TeV through the process $e^{+}e^{-}
\rightarrow e^{+}\gamma^{*} \gamma^{*} e^{-} \rightarrow e^{+}\, Z\,
Z\, e^{-}$.}
\label{Fig.18}
\end{figure}

The SM cross section of the processes $\gamma\gamma\rightarrow ZZ$, $e^{-}\gamma\rightarrow
e^{-}\gamma^{*}\gamma \rightarrow e^{-}Z\, Z$ and $e^{+}e^{-}
\rightarrow e^{+}\gamma^{*} \gamma^{*} e^{-} \rightarrow e^{+}\, Z\,
Z\, e^{-}$ is quite small, because the process $\gamma\gamma\rightarrow ZZ$ and the subprocesses $\gamma^*\gamma \to ZZ$ and $\gamma^*\gamma^* \to ZZ$ are not allowed at the tree level. They are only allowed at loop level and can be neglected. On the other hand, as stated in Ref. \cite{Diakonidis:2006cv},  the SM background and
their interference contributions of the examined processes may be important for low center-of-mass energies such as 0.35 TeV and 1.4 TeV. However, the effect of the one-loop SM cross section at $\sqrt{s}=3$ TeV of these processes is expected to give relatively small contributions and it can be neglected. For this reason, we analysis  anomalous $ZZ\gamma\gamma$  quartic couplings only at $\sqrt{s}=3$ TeV for three processes. Therefore, in the course of statistical analysis, the bounds of all anomalous quartic couplings at 95 \% C.L. are calculated using the Poisson statistics test since the number of SM background events of the examined processes expected to be negligible events for the various values of the luminosities at $\sqrt{s}=3$ TeV. In this case, the upper
bounds of number of events $N_{up}$ at the 95 \% C.L. can be calculated from the following formula
\begin{eqnarray}
\sum_{k=0}^{N_{obs}} P_{Poisson}(N_{up}; k)=1-CL.
\end{eqnarray}
where $N_{obs}$ is the number of observed events and the value of $N_{up}$ can be obtained with respect to the value of the number of observed events. For calculating the limits on anomalous quartic gauge couplings in case there is no signal, $N_{obs}$=0, and then $N_{up}$ is always 3, for 95 \% C.L. This upper limit on the number of events is translated, in each case separately, to an upper/lower limit on the anomalous quartic gauge couplings, using the cross-section dependence on the anomalous quartic gauge couplings at the corresponding energy, and multiplying the cross section by the branching ratio for leptonic Z decays and by the corresponding luminosity. The bounds at 95\% C.L. on these couplings at the CLIC with $\sqrt s$=3 TeV for various integrated luminosities are shown in Figs. \ref{Fig.10}-\ref{Fig.18} for the examined processes. Here we consider that only one of the anomalous couplings changes at any time. 
 
 As can be seen from Fig. \ref{Fig.10}, the sensitivity bounds of $f_{M2}/\Lambda^4$ and $f_{M3}/\Lambda^4$ couplings obtained from the process $\gamma\gamma \to ZZ$ with $\sqrt s$= 3 TeV and $L_{int}=2000$ fb$^{-1}$ are calculated as $[-3.30;3.30]\times 10 ^{-3}$ TeV$^{-4}$ and $[-1.20;1.20]\times 10 ^{-2}$ TeV$^{-4}$ which are seven and six orders of magnitude better than the experimental bounds of the LHC, respectively. The expected best sensitivities on $f_{T0}/\Lambda^4$ and $f_{T9}/\Lambda^4$ couplings in Fig. \ref{Fig.10} are far beyond the sensitivities of the LHC.   As can be seen from Table \ref{tab4}, when the luminosity reduction factor is taken into account, these limits become $[-8.45;8.45]\times 10 ^{-3}$ TeV$^{-4}$  and $[-3.17;3.17]\times 10 ^{-2}$ TeV$^{-4}$, respectively. So, the sensitivity of the limits calculated using luminosity reduction factors decrease by about 2.5 times for $\gamma\gamma$ option and 1.6 times for $e\gamma$ option in $e^+e^-$ collisions. 
 
We compare our results with the best bounds obtained from the phenomenological studies of the LHC, future hadron and linear colliders in the literature. The bounds on $\frac{a_{0}}{\Lambda^{2}}$ and $\frac{a_{c}}{\Lambda^{2}}$ couplings arising from dimension-6 operators have been obtained by Refs. \cite{Chapon:2009hh, N.Cartiglia:2015gve}. For 95$\%$ C. L. with integrated luminosity of 200 fb$^{-1}$ at $\sqrt{s}=14$ TeV at the LHC, the sensitivities on the anomalous couplings are calculated as $[-1.1; 1.1]$ TeV$^{-2}$ and $[-4.8; 4.8]$ TeV$^{-2}$, respectively. However, the best sensitivities on $\frac{a_{0}}{\Lambda^{2}}$ and $\frac{a_{c}}{\Lambda^{2}}$ couplings for $L_{int}=590$ fb$^{-1}$ at $\sqrt{s}=3$ TeV at the CLIC are at the order of 10$^{-2}$ TeV$^{-2}$ \cite{murat}. Also, Refs. \cite{Degrande:2013yda,Baak:2013fwa} have investigated the couplings of dimension-8 operators at 95$\%$ C. L. with integrated luminosity of 300 fb$^{-1}$ at $\sqrt{s}=14$ TeV and 3000 fb$^{-1}$ at $\sqrt{s}=14$ TeV, 33 TeV and 100 TeV at the LHC and future hadron colliders. The bounds on the couplings arising from dimension-8 operators are given $\frac{f_{M2}}{\Lambda^{4}}$=25 TeV $^{-4}$ and $\frac{f_{M3}}{\Lambda^{4}}$=38 TeV$^{-4}$.

In Table \ref{tab4}, we show the best sensitivity bounds at 95$\%$ C. L. of $\frac{f_{M2,3}}{\Lambda^{4}}$ and and $\frac{a_{0,c}}{\Lambda^{2}}$ couplings for three processes with integrated luminosity 2000 fb$^{-1}$ at $\sqrt{s}=3$ TeV. As can be seen in Table \ref{tab4}, our best sensitivities on $\frac{a_{0,c}}{\Lambda^{2}}$ couplings by examining the process $\gamma\gamma\rightarrow ZZ$ are about 10$^{5}$ times better than the sensitivities calculated in Refs. \cite{Chapon:2009hh, N.Cartiglia:2015gve}. Our bounds can set more stringent sensitivity by three orders of magnitude with respect to the best sensitivity derived from the CLIC with $\sqrt{s}=3$ TeV. Finally, we can understand from Table IV that the best bounds obtained through the process $\gamma\gamma \rightarrow Z Z$ with integrated luminosity 2000 fb$^{-1}$ at $\sqrt{s}=3$ TeV improve the sensitivities of $\frac{f_{M2}}{\Lambda^{4}}$ and $\frac{f_{M3}}{\Lambda^{4}}$ couplings by up to a factor of 10$^{4}$
compared to Refs. \cite{Degrande:2013yda,Baak:2013fwa}. However, we compare
our results with the sensitivities of Ref. \cite{Degrande:2013yda} which investigates phenomenologically $f_{T9}/\Lambda^4$ coupling via $p p\to ZZ+ 2j \to 4 l+ 2j$ process at $\sqrt s$= 14 (33) TeV with 300 (3000) fb$^{-1}$ luminosity. The bound on $f_{T9}/\Lambda^4$ coupling at $\sqrt s$= 33 TeV with $L_{int}=3000$ fb$^{-1}$ is [$-2.50$; $2.50$] TeV$^{-4}$ which is up to a factor of 10$^{3}$ worse than our best bound.
However, it can be seen from Fig. \ref{Fig.14} that bounds on $\frac{f_{T9}}{\Lambda^{4}}$ coupling obtained
from the process $e^{-}\gamma\rightarrow e^{-}\gamma^{*}\gamma \rightarrow e^{-}Z\, Z$ are more restrictive than the bounds on $\frac{f_{M2}}{\Lambda^{4}}$, $\frac{f_{M3}}{\Lambda^4}$ and $\frac{f_{T0}}{\Lambda^4}$ couplings. 
The best sensitivities obtained for four different couplings from the process $\gamma\gamma\rightarrow ZZ$ in Fig. \ref{Fig.10} are approximately an order of magnitude more restrictive with respect to the main process $e^{+}e^{-} \rightarrow e^{+}\gamma^{*} \gamma^{*} e^{-} \rightarrow e^{+}\, Z\,Z\, e^{-}$ in Fig. \ref{Fig.18} which is obtained by integrating the cross section for the subprocess $\gamma^*\gamma^* \to ZZ$ over the effective photon luminosity. Although the luminosity reduction factor is taken into account in $\gamma\gamma$ and $e\gamma$ collision modes, the results show that $\gamma\gamma$ collisions give the best bounds to test anomalous quartic gauge couplings with respect to $\gamma^*\gamma^*$ and $\gamma \gamma^*$ collisions. Principally, the sensitivity of the processes to anomalous couplings rapidly increases with the center-of-mass energy and the luminosity.

\section{Conclusions}
CLIC is envisaged as a high energy $e^+e^-$ collider having with very clean experimental conditions and being free from strong interactions with respect to the LHC. In addition, the number of SM events vanishes for $\gamma\gamma\rightarrow ZZ$, $e^{-}\gamma\rightarrow e^{-}\gamma^{*}\gamma \rightarrow e^{-}Z\, Z$ and $e^{+}e^{-} \rightarrow e^{+}\gamma^{*} \gamma^{*} e^{-} \rightarrow e^{+}\, Z\,Z\, e^{-}$ processes. Therefore, the observation of a few events at the final state of such processes would be an important sign for anomalous quartic couplings beyond the SM.
For these reasons, we have estimated the improvement of sensitivity to anomalous quartic $ZZ\gamma\gamma$ couplings with dimension-8 as function of collider energies and luminosities through the processes $\gamma\gamma\rightarrow ZZ$, $e^{-}\gamma\rightarrow e^{-}\gamma^{*}\gamma \rightarrow e^{-}Z\, Z$ and $e^{+}e^{-} \rightarrow e^{+}\gamma^{*} \gamma^{*} e^{-} \rightarrow e^{+}\, Z\,Z\, e^{-}$.  As a result, the CLIC as photon-photon collider provides an ideal platform to examine anomalous quartic $ZZ\gamma\gamma$ gauge couplings at high energies and luminosities.



\pagebreak

\begin{table}[htbp]
\caption{The total cross sections as function of anomalous $f_{M2}/\Lambda^4$, $f_{M3}/\Lambda^4$, $f_{T0}/\Lambda^4$ and $f_{T9}/\Lambda^4$ couplings assuming $\Lambda$ =1 TeV for the processes $\gamma\gamma \to ZZ$, $e^{-}\gamma\rightarrow
e^{-}\gamma^{*}\gamma \rightarrow e^{-}Z\, Z$ and $e^{+}e^{-}
\rightarrow e^{+}\gamma^{*} \gamma^{*} e^{-} \rightarrow e^{+}\, Z\,
Z\, e^{-}$ at CLIC with $\sqrt s=3$ TeV.}
\begin{center}
\begin{tabular}{c ccccc c c c} \hline \hline
Modes &&&&Total cross sections (pb)&&&&\\ \hline \hline
$\gamma\gamma$ && 31.0$f_{M2}^2$ && 2.21$f_{M3}^2$  &&  28.36$f_{T0}^2$ &&103.5  $f_{T9}^2$ \\ 
$e\gamma$ &&1.96$f_{M2}^2$ && 0.14$f_{M3}^2$ &&  1.80$f_{T0}^2$ &&6.60$f_{T9}^2$ \\
$e^+e^-$&& $1.24\times 10^{-1}$$f_{M2}^2$ && $8.86\times 10^{-3}$$f_{M3}^2$  &&  $1.14\times 10^{-1}$$f_{T0}^2$ &&$4.19\times 10^{-1}$$f_{T9}^2$  \\ \hline\hline
\end{tabular}
\end{center}
\label{tab1}
\end{table}

\begin{table}[htbp]
\caption{$95\%$ C.L. sensitivity bounds of $a_{0}/\Lambda^2$, $a_{c}/\Lambda^2$, $f_{M2}/\Lambda^4$, $f_{M3}/\Lambda^4$, $f_{T0}/\Lambda^4$ and $f_{T9}/\Lambda^4$ couplings for 2000 fb$^{-1}$ integrated luminosity and $\sqrt s$=3 TeV through the processes $\gamma\gamma \to ZZ$, $e^{-}\gamma\rightarrow
e^{-}\gamma^{*}\gamma \rightarrow e^{-}Z\, Z$ and $e^{+}e^{-}
\rightarrow e^{+}\gamma^{*} \gamma^{*} e^{-} \rightarrow e^{+}\, Z\,
Z\, e^{-}$. The values in the parentheses are calculated limits using the luminosity reduction factor. }
\begin{center}
\begin{tabular}{c ccccc c c} \hline \hline
Couplings && $\gamma\gamma$ mode&&$e \gamma$ mode && $e^+e^-$ mode \\ \hline \hline
$a_{0}/\Lambda^2$ (TeV$^{-2}$) &&[-4.27;4.27]$\times 10^{-5}$   && [-1.68;1.68]$\times 10^{-4}$    && [-6.68;6.68]$\times 10^{-4}$     \\ 
                                   &&([-1.09;1.09]$\times 10^{-4}$) &&([-2.69;2.69]$\times 10^{-4}$)     &&     \\
$a_{c}/\Lambda^2$ (TeV$^{-2}$) &&[-1.55;1.55]$\times 10^{-4}$   && [-6.29;6.29]$\times 10^{-4}$    && [-2.46;2.46]$\times 10^{-3}$     \\ 
                                   &&([-4.11;4.11]$\times 10^{-4}$) &&([-1.01;1.01]$\times 10^{-3}$)     &&     \\
$f_{M2}/\Lambda^4$  (TeV$^{-4}$)  &&[-3.30;3.30]$\times 10^{-3}$   && [-1.30;1.30]$\times 10^{-2}$    && [-5.16;5.16]$\times 10^{-2}$     \\ 
       &&([-8.45;8.45]$\times 10^{-3}$) &&([-2.08;2.08]$\times 10^{-2}$)     &&     \\
$f_{M3}/\Lambda^4$  (TeV$^{-4}$)  &&[-1.20;1.20]$\times 10^{-2}$   && [-4.86;4.86]$\times 10^{-2}$   &&[-1.90;1.90]$\times 10^{-1}$    \\ 
                      &&([-3.17;3.17]$\times 10^{-2}$) &&([-7.79;7.79]$\times 10^{-2}$)      &&   \\
$f_{T0}/\Lambda^4$ (TeV$^{-4}$)  &&[-3.40;3.40]$\times 10^{-3}$  &&[-1.36;1.36]$\times 10^{-2}$   &&[-5.39;5.39]$\times 10^{-2}$     \\
                      &&([-8.83;8.83]$\times 10^{-3}$) && ([-2.17;2.17]$\times 10^{-2}$)   &&    \\
$f_{T9}/\Lambda^4$ (TeV$^{-4}$)  &&[-1.80;1.80]$\times 10^{-3}$  && [-7.09;7.09]$\times 10^{-3}$ && [-2.81;2.81]$\times 10^{-2}$    \\
                        &&([-8.83;8.83]$\times 10^{-3}$)&&([-1.14;1.14]$\times 10^{-2}$) 	&&\\
 \hline\hline
\end{tabular}
\end{center}
\label{tab4}
\end{table}

\end{document}